\documentclass[12pt]{article}
\usepackage{amssymb}
\def\beq{\begin{equation}}
\def\eeq{\end{equation}}
\usepackage[all,2cell,dvips]{xy}
\def\IR{\relax{\rm I\kern -.18em R}}

\begin{document}
\title{ Supersymmetric Exact Sequence, Heat Kernel and Super KdV Hierarchy }
\author{ \Large S. Andrea*, A. Restuccia**, A. Sotomayor***}
\maketitle{\centerline {*Departamento de Matem\'{a}ticas,}}
\maketitle{\centerline{**Departamento de F\'{\i}sica}}
\maketitle{\centerline{Universidad Sim\'on Bol\'{\i}var}}
\maketitle{\centerline{***Departamento de Ciencias B\'{a}sicas}}
\maketitle{\centerline{Unexpo, Luis Caballero Mej\'{\i}as }}
\maketitle{\centerline{e-mail: sandrea@usb.ve, arestu@usb.ve,
asotomay@zeus.unexpo.edu.ve }}
\begin{abstract}We introduce the free $N=1$ supersymmetric
derivation ring and prove the existence of an exact sequence of
supersymmetric rings and linear transformations. We apply
necessary and sufficient conditions arising from this exact
supersymmetric sequence to obtain the essential relations between
conserved quantities, gradients and the $N=1$ super KdV hierarchy.
We combine this algebraic approach with an analytic analysis of
the super heat operator. We obtain the explicit expression for
the Green's function of the super heat operator in terms of a
series expansion and discuss its properties. The expansion is
convergent under the assumption of bounded bosonic and fermionic
potentials. We show that the asymptotic expansion when
$t\rightarrow0^+$ of the Green's function for the super heat
operator evaluated over its diagonal generates all the members of
the $N=1$ super KdV hierarchy.

\end{abstract}

\section{Introduction}The analysis of supersymmetric quantum
problems has been recently considered in several relevant physical
contexts. At very high energies one way of studying the
$M$-theory, which has been proposed as a theory of unification of
all known interactions in nature, is through Matrix models
describing supersymmetric quantum problems. The supersymmetry is
one of the relevant ingredients of these models. In particular the
presence of supersymmetry may change completely the spectrum of
the quantum hamiltonian. The bosonic hamiltonian in the case of
the $D=11$ supermembrane, with Minkowski target space, has a
discrete spectrum while its supersymmetric extension has a
continuous spectrum. Moreover in that case the Green's function
does not admit a representation in terms of the Feynman path
integral since the potential is not bounded from below in some
directions in spite of the fact that the SUSY hamiltonian is
nonnegative. These systems are very closely related to certain
supersymmetric integrable systems. These models in themselves are
a great help in understanding integrable systems. The $N=1,2$
supersymmetric extensions of the KdV equations were found several
years ago  \cite{M,Manin,Laberge,Popowicz} , while the $N=3$ and
$N=4$ have been considered more recently in \cite{Yung,Krivonos,
Delduc}. The bi-hamiltonian structure of the super KdV equations
was studied in \cite{Oevel,Farrill}. For a review of $N=1$ and
$N=2$ super KdV equations see \cite{Mathieus}. Extensions of the
supersymmetric model have been proposed in \cite{Andrea}.

In the first part of this work we focus on the algebraic structure
of $N=1$ supersymmetric models. We introduce the ``Free SUSY
derivation ring on a single fermionic generator" which contains a
parity automorphism, a canonical superderivation and a
supersymmetric gradient operator. We established an exact sequence
of SUSY rings and linear transformations. Several of these
relations were already known in the literature \cite{Manin,MM}.
This is a general algebraic construction valid for one dimensional
supersymmetric models. Using necessary and sufficient conditions
arising from the exact sequence we obtain the essential relations
between conserved quantities, gradients and the $N=1$ SKdV
hierarchy. In the second part of this work we use the exact
sequence together with an analytic analysis of the super heat
kernel to show that the asymptotic expansion when
$t\rightarrow0^+$ of the Green's function for the super heat
operator, evaluated over its diagonal, generates all the left
hand members of the SKdV hierarchy. Using the symmetry properties
of that Green's function we obtain an iterative procedure to
obtain all the gradients and members of the SKdV hierarchy. The
algebraic approach arising from the exact SUSY sequence together
with the analytic approach arising from the analysis of the
Green's function of the superheat operator combine to give a
satisfactory description of the SKdV hierarchy.

\section{The Free Supersymmetric Derivation Ring on a single
Fermionic Generator}
\subsection{Definitions} A ring $\mathcal{A}$ is associative but not
necesarily commutative. It is ``oriented" when there is given a
ring automorphism $P:\mathcal{A}\rightarrow \mathcal{A}$
satisfying $P^2=I$, as well as $P(f+g)=Pf+Pg$ and
$P(fg)=(Pf)(Pg)$. An element $f$ of $\mathcal{A}$ is ``oriented"
when $Pf=\pm f$, $f$ being called ``bosonic" when $Pf=f$ and
``fermionic" when $Pf=-f$. The oriented ring $(\mathcal{A},P)$ is
said to be ``commutative" when
\[fg=gf(-1)^{\sigma\tau}\] wherever $f,g\in \mathcal{A}$ satisfy
$Pf=(-1)^\sigma f$, $Pg=(-1)^\tau g$. In this situation the
commutation formula \[uf=f(P^\sigma u)\] holds for all $u\in
\mathcal{A}$, given that $Pf=(-1)^\sigma f$; this follows from
$u=u_0+u_1=boson+fermion$. A linear map $D:\mathcal{A}\rightarrow
\mathcal{A}$ is called an ``ordinary derivation" when
\begin{eqnarray*}& &DP=PD
\\ & &D(uv)=(Du)v+u(Dv)\end{eqnarray*} for all $u,v\in \mathcal{A}$, and is called a
``superderivation" when
\begin{eqnarray*} & &DP = -PD \\ & &D(uv) = (Du)v+(Pu)(Dv). \end{eqnarray*} The two definitions may be
combined by saying that $\Pi(D)=(-1)^\delta$ when
\begin{eqnarray*}& &DP=(-1)^\delta
PD\\ & &D(uv)=(Du)v+(P^\delta u)Dv.\end{eqnarray*} Then, if
$(\mathcal{A},P)$ is commutative and $f$ satisfies $Pf=(-1)^\sigma
f$, the product $fD$ will also be a derivation, with
\[\Pi(fD)=(-1)^{\sigma+\delta}.\] Thus a fermionic $f$ times a
bosonic $D$ will be a fermionic $fD$, and similarly for the other
three cases.
\subsection{Construction of the ring $\mathcal{A}$} We may begin with the
commutative ring $\mathcal{B}$ of all polynomials in the commuting
letters $b_2,b_4,b_6...$  The ordinary derivations
$\frac{\partial}{\partial b_n}:\mathcal{B\rightarrow}\mathcal{B}$
all commute, as $n$ runs through even positive integers. Then
$\mathcal{A}$ is the supersymmetric extension of $\mathcal{B}$,
constructed as follows. Let $ \mathbb{M}=\left\{1,3,5,...\right\}$
be the set of all positive odd numbers, and let
$2^\mathbb{M}=\left\{\phi,1,3,13,...\right\}$ be the collection of
all finite subsets of $\mathbb{M}$, including the empty set
$\phi$. Then $\mathcal{A}$ is to consist of all finitely supported
functions $f:2^\mathbb{M}\rightarrow \mathcal{B}$. The product of
two elements $f$ and $g$, evaluated at a finite subset
$E\subset\mathbb{M}$, is defined to be \[(fg)(E)=\sum_{A\bigcup
B=E}f(A)g(B)\varepsilon(A,B).\] Here the function
$\varepsilon:2^\mathbb{M}\times
2^\mathbb{M}\rightarrow\left\{-1,0,1\right\}$ is defined to be
\[\varepsilon(A,B)=\prod_{a\in A}\prod_{b\in B}\varepsilon(a,b),\] where
$\varepsilon:\mathbb{M}\times\mathbb{M}\rightarrow
\left\{-1,0,1\right\}$ is defined by
\[\varepsilon\left(a,b\right)= \left\{\begin{array}{cc} 1 & a<b \\
0 & a=b \\ -1 & a>b
\end{array}\right.\] If the number of elements in $E$ is $|E|$,
then the above sum has $2^{|E|}$ terms, since
$\varepsilon(A,B)\neq 0$ only occurs when $A$ and $B$ are
disjoint. The parity automorphism $P:\mathcal{A}\rightarrow
\mathcal{A}$ is given by $(Pf)(E)=f(E)(-1)^{|E|}.$

The easy formula
$\varepsilon\left(A,B\right)=\varepsilon\left(B,A\right)\left(-1\right)^{|A|
|B| }$ makes it clear that $(\mathcal{A},P)$ is commutative.

When $B,C\subset \mathbb{M}$ are disjoint, one has
$\varepsilon(A,B\cup C)=\varepsilon(A,B)\varepsilon(A,C).$ This
shows that the product operation is associative, the value of
$\left(fg\right)h=f\left(gh\right)$ on $E\subset \mathbb{M}$ being
given by \[\sum_{A\cup B\cup
C=E}f(A)g(B)h(C)\varepsilon(A,B)\varepsilon(A,C)\varepsilon(B,C).\]

This completes the construction of the oriented ring
$\left(\mathcal{A},P\right).$ The generating elements
$a_1,a_2,a_3,...\in \mathcal{A}$ are now to be identified.

The inclusion $\mathcal{B}\subset \mathcal{A}$ is realized by
associating each $b\in \mathcal{B}$ with that function
$2^\mathbb{M}\rightarrow \mathcal{B}$ wich sends the empty set
$\phi$ to $b$, and everything else to zero. Thus, for $m$ even,
$a_m(\phi)=b_m$ and $a_m(E)=0$ for all nonempty $E\subset
\mathbb{M}.$

When $p$ is an odd positive integer,
$a_p:2^\mathbb{M}\rightarrow\mathcal{B}$ is defined by
$a_p(p)=1\in \mathcal{B}$ and $a_p(E)=0$ for all other finite
subsets of $\mathbb{M}.$

This gives us $\left\{a_1,a_2,a_3,...\right\}\subset \mathcal{A}.$

Evidently $Pa_n=(-1)^na_n.$ Further, if $1\leq p_1<p_2<...<p_n$
are odd, the product $a_{p_1} a_{p_2}\cdots a_{p_m}\in
\mathcal{A}$ takes the value $+1$ on the subset
$\left\{p_1,p_2,...,p_n\right\}\subset \mathbb{M}$ and zero
everywhere else. Therefore every element of $\mathcal{A}$ may be
written as a finite polynomial in the elements $a_1,a_2,a_3,...$

\subsection{Derivations of $\mathcal{A}$}

The fundamental superderivation of $\mathcal{A}$ is defined by
\[D=a_2\frac{\partial}{\partial a_1}+a_3\frac{\partial}{\partial
a_2}+ a_4\frac{\partial}{\partial a_3}+\cdots\] It sends
$a_1\rightarrow a_2\rightarrow a_3\rightarrow\cdots$, exchanging
bosons and fermions. This suggests that $(\mathcal{A},P,D)$ is in
some sense the natural model of an oriented superderivation ring
generated by a single fermion $a_1$.

The square of a superderivation is an ordinary derivation. Thus
$D^2$ restricted to $\mathcal{B}\subset \mathcal{A}$ sends
$a_2\rightarrow a_4\rightarrow a_6\cdots$, and the pair
$(\mathcal{B},D^2)$ can be called the free bosonic derivation ring
on a single generator.

The inclusions $D\mathcal{A}_n\subset \mathcal{A}_n$ which follow
from $DE=ED$ will be used later in the proofs of the exact
sequence.

\subsection{The Supersymmetric Gradient Operator} Given $h\in
\mathcal{A}$ we ask whether $f\subset \mathcal{A}$ exists with
$h=Df$. A linear operator $M:\mathcal{A}\rightarrow \mathcal{A}$
with $MD=0$ would give at least a necessary condition.

In analogy with the bosonic case, for which it is known the
gradient operator $M$, it can be found that the Susy $M$ operator
has the expression
\[M=\frac{\partial}{\partial a_1}+D\frac{\partial}{\partial
a_2}-D^2\frac{\partial}{\partial a_3}-D^3\frac{\partial}{\partial
a_4}+D^4\frac{\partial}{\partial a_5}+\cdots\] It is a linear
operator sending $\mathcal{A}$ into itself, and we have seen that
the equation $Mh=0$ is a necessary condition for the existence of
$f\in \mathcal{A}$ with $Df=h.$

Later it will be shown that the condition $Mh=0$ is also
sufficient. (With respect to parity we note that $PM=-MP$).

\subsection{Operators and Adjoints} Given $(\mathcal{A},P,D)$
as constructed: a ``differential operator" is a linear map of
$\mathcal{A}$ into itself having the form \[L=\sum_0^Nl_nD^n\]
with coefficients $l_n\in \mathcal{A}$. $L$ is the identically
zero map $\mathcal{A}\rightarrow \mathcal{A}$ if and only if all
the coefficients are zero.

Then $\mathcal{O}_p\mathcal{A}$ is defined to be the set of all
differential operators. It is an associative ring: the composition
of two operators has the same form, because $D_n(fI)$ can be
expanded as a finite sum $\sum_{r=0}^ng_rD^r$, by using the
defining property of $D$ on products of elements of $\mathcal{A}$.

Composing $L$ with the parity automorphism
$P:\mathcal{A}\rightarrow \mathcal{A}$ we find that
\[PLP=\sum_0^N(Pl_n)(-D)^n.\]

Therefore $O_p\mathcal{A}$ is also an oriented ring, its parity
automorphism given by $L\rightarrow PLP.$ As before, $L$ can be
called ``oriented" if $PL=\pm LP,$ ``fermionic" in one case and
``bosonic" in the other.

We now ask how to integrate by parts in $\mathcal{A}.$ Suppose
$u,v\in \mathcal{A}$ are oriented elements with
$Pu=u(-1)^\alpha,Pv=v(-1)^\beta.$ Starting with $D\in
\mathcal{O}_p,$ we compute \begin{eqnarray*}
D(uv)&=&(Du)v+(-1)^\alpha u(Dv) \\
&=&(Du)v+(-1)^\alpha(Dv)u(-1)^{\alpha(\beta+1)} \\
&=&(Du)v+(Dv)u(-1)^{\alpha \beta}.\end{eqnarray*}

This can be written $(Du)v\equiv (-Dv)u(-1)^{\alpha \beta},$ if
the congruence notation $f\equiv g$ in $\mathcal{A}$ is defined to
mean that $f-g=Dh$ for some $h\in \mathcal{A}.$

More generally, $L$ and $L^*\in \mathcal{O}_p\mathcal{A}$ may be
said to be ``mutually adjoint" if \[(Lu)v\equiv
(L^*v)u(-1)^{\alpha \beta}\] for all oriented $u,v$ as above.
Thus, $D^*=-D$, while a zeroth order operator $l_0I$ is its own
adjoint.

The uniqueness of the adjoint operator is argued as follows: if
$L=0$ then every $h=L^*v$ satisfies $hu\equiv 0$ for all $u\in
\mathcal{A}.$ However, it can be shown that for any nonzero $h\in
\mathcal{A}$ there exist $u$ such that $hu$ cannot be of the form
$Df$ for any $f\in \mathcal{A}.$

Consequently, $L=0$ in $\mathcal{O}_p\mathcal{A}$ implies that all
$L^*v=0$ in $\mathcal{A},$ and hence $L^*=0$ in
$\mathcal{O}_p\mathcal{A}.$ This shows that any $L\in
\mathcal{O}_p\mathcal{A}$ can have at most one adjoint $L^*\in
\mathcal{O}_p\mathcal{A},$ and furthermore that $(L^*)^*=L.$

The commutation of the constructions $L\rightarrow PLP$ and
$L\rightarrow L^*$ is shown by applying $P$ to the
congruence\[(LPu)Pv\equiv (L^*Pv)Pu(-1)^{\alpha \beta}.\]

Thus if $L$ has an adjoint then so does $PLP$, and
\[(PLP)^*=PL^*P.\]

The existence of adjoints for all differential operators must now
be shown.

$\smallskip$

\noindent \textbf{Proposition} Suppose $K,L\in
\mathcal{O}_p\mathcal{A}$ have adjoints, and that
$PKP=K(-1)^\kappa, PLP=L(-1)^\lambda .$ Then $KL$ has an adjoint,
and it is given by
\[(KL)^*=L^*K^*(-1)^{\kappa \lambda}.\]

$\smallskip$

The proposition generalizes inmediately to finite products of
operators $L_1L_2\cdots L_m$ in which each $L_k$ has an adjoint
and is oriented with $PL_kP=L_k(-1)^{\lambda_k}.$ Then
\[(L_1L_2\cdots L_m)^*=(L_m^*L_{m-1}^*\cdots L_1^*)(-1)^\mu,\]
\[\mu=\sum_{1\leq i<j\leq m}\lambda_i\lambda_j.\] Thus $lD^k$,
when $Pl=\pm l,$ has an adjoint $\pm D^k(lI),$ the sign depending
on $k$ and the parity of $l$. This proves that every $L\in
\mathcal{O}_p\mathcal{A}$ possesses a unique adjoint $L^*\in
\mathcal{O}_p\mathcal{A}$, the bijection $L\leftrightarrow L^*$
satisfying $(L^*)^*=L.$

We conclude by computing the adjoint of $D^plD^q,p+q=m.$ If
$Pl=-l$ then all the $m+1$ exponents are $+1$ and
$\mu=\frac{m(m+1)}{2}.$

Then
\begin{eqnarray*}(D^plD^q)^*&=&(-D)^ql(-D)^p(-1)^\frac{m(m+1)}{2}
\\ &=&D^qlD^p(-1)^{m+\frac{m(m+1)}{2}} \\
&=&D^q(P^ml)D^p(-1)^{\frac{m(m+1)}{2}}.\end{eqnarray*}

On the other hand, if $Pl=+l$ then all but one of the exponents
$\lambda_1,...,\lambda_{m+1}$ is $+1$, the remaining exponent
being zero. Then
\begin{eqnarray*}(D^plD^q)^*&=&(-D)^ql(-D)^p(-1)^{\frac{m(m-1)}{2}}
\\ &=&D^qlD^p(-1)^{\frac{m(m+1)}{2}} \\
&=&D^q(P^ml)D^p(-1)^{\frac{m(m+1)}{2}}.\end{eqnarray*}

But every $l\in \mathcal{A}$ is uniquely the sum of a boson and a
fermion. Therefore, for any $l\in \mathcal{A}$ and any nonnegative
integers $p$ and $q$, the adjoint of $D^plD^q\in
\mathcal{O}_p\mathcal{A}$ is given by
\[(D^plD^q)^*=D^q(P^ml)D^p(-1)^{\frac{m(m+1)}{2}}\], where $p+q=m$.

\subsection{Frechet derivative} The construction of the Frechet
derivative operator gives a linear map $\mathcal{A}\rightarrow
\mathcal{A}$, sending $f\rightarrow L_f.$ Given
$f(a_1,a_2,...,a_n)$ an element of $\mathcal{A}$, the action of
$L_f$ on a fermionic element $v\in \mathcal{A}, Pv=-v,$ may be
defined by
\[L_fv=\frac{d}{d\epsilon}|^{\epsilon=0}f(a_1+\epsilon v,a_2+\epsilon Dv,...,a_n+\epsilon D^{n-1}v).\]

The coefficients of $L_f$ are obtained as follows. If $q$ is odd
and $f=ga_qh$ with $g$ and $h$ independent of $a_q$, we have
$\frac{\partial}{\partial a_q}f=(Pg)h$, while
$g(D^{q-1}v)h=g(Ph)(D^{q-1}v)$ since $D^{q-1}v$ is fermionic.
Therefore $D^{q-1}v$ is multiplied on the left by
$(P\frac{\partial}{\partial a_q}f).$ If $m$ is even then
$\frac{\partial}{\partial a_m}$ is an ordinary derivation and
$D^{m-1}v$ is bosonic. In this case there is no anticommutation,
and $D^{m-1}v$ is multiplied on the left by $\frac{\partial
f}{\partial a_m}.$ Combining these two cases we obtain the general
formula \[L_f=\sum_{n=1}^\infty (P^n\frac{\partial f}{\partial
a_n})D^{n-1},\] which gives the Frechet derivative operator
$L_f\in \mathcal{O}_p\mathcal{A}$ for any $f\in \mathcal{A}.$

Applying $L_f$ to the generating element $a_1\in \mathcal{A}$ we
get \[L_fa_1=\sum_{n=1}^\infty(P^n\frac{\partial f}{\partial
a_n})a_n.\] But $ha_n=a_n(P^nh)$ for all $h\in \mathcal{A}.$ Hence
$L_fa_1=Ef,$ connecting $L_f$ to the Euler operator $E.$

The adjoint operator to $L_f$ may be written down using the
results of the previous section:
\[{L_f}^*=\sum_{n=1}^\infty(-1)^{\frac{n(n-1)}{2}}D^{n-1}((P\frac{\partial
f}{\partial a_n}I)).\]

Applying the operator parity automorphism $K\rightarrow PKP$ we
get
\[P{L_f}^*P=\sum_{n=1}^\infty(-1){\frac{n(n-1)}{2}}(-D)^{n-1}(\frac{\partial f}{\partial
a_n}I).\]

The coefficient of the identity operator is readily accessible
since\[D^{n-1}(\frac{\partial f}{\partial
a_n}I)=(D^{n-1}\frac{\partial f}{\partial a_n})I+(?)D+\cdots\]

After checking the $\pm$ signs we see that this coefficient is
just the supersymmetric gradient $Mf$ of $f$: thus
\[{L_f}^*=(PMf)I+(?)D+\cdots\]

Going back to the equation $Mh=0$, we see that this condition
would imply the existence of $K\in \mathcal{O}_p\mathcal{A}$
satisfyind ${L_h}^*=KD$ and hence also $L_h=\pm DK^*$ if h is
oriented. Applying this operator equation to the generating
element $a_1\in \mathcal{A}$, we find that $Eh=Df$ for some $f\in
\mathcal{A}.$

Thus $Mh=0$ implies $h\equiv 0$ , at least when $h$ is oriented
and $Eh=nh,n> 0$. But the two extra conditions are no obstacle:

$\smallskip$

\noindent \textbf{Proposition} Given $h\in \mathcal{A}$, with zero
constant term. Then $Mh=0$ if and only if $h=Df$ for some $f\in
\mathcal{A}.$

 \noindent \textbf{Proof} The equation $PM=-MP$ shows
that $h\pm Ph$ also has zero gradient. Replacing $h$ by either
summand in $h=boson+fermion$, we may suppose that $h$ is oriented.
The presentation
\[\mathcal{A}=\mathcal{A}_0\oplus\mathcal{A}_1\oplus\mathcal{A}_2\oplus\cdots\]

by eigenspaces of the Euler operator permits $h$ to be written as
$h=h_0+h_1+h_2+\cdots,$ in which $h_0=0$ by hypothesis. Then,
since $M\mathcal{A}_n\subset\mathcal{A}_{n-1}$ the condition
$Mh=0$ implies $Mh_n=0$ for all $n\geq 1.$

By what was said earlier, there exist $f_n$ with $Df_n=Eh_n=nh_n.$
Therefore $h=D(f_1+\frac{1}{2}f_2+\frac{1}{3}f_3+\cdots)$ and the
proof is complete.

$\smallskip$

In the next section we will need to know the interaction between
the $f\rightarrow L_f$ construction and the parity automorphism
$P:\mathcal{A}\rightarrow \mathcal{A}.$ The formula
$L_{Pf}=-PL_fP$ is easily verified by
\begin{eqnarray*}(P^n\frac{\partial}{\partial
a_n}Pf)D^{n-1}&=&(-1)^n\left(P^{n+1}\frac{\partial f}{\partial
a_n}\right)D^{n-1} \\&=&-\left(P(P^n\frac{\partial f}{\partial
a_n})\right)(-D)^{n-1}.\end{eqnarray*}

\subsection{Gradients and operators} We wish to determine which
elements $g\in \mathcal{A}$ are of the form $Mh$ for some $h\in
\mathcal{A}$: in the applications this asks which $g$ are
gradients of possible conserved quantities. Here are two facts,
that can be proven easily:

(i) For any $f\in \mathcal{A}$, the Frechet derivative operators
of $f$ and of $Df$ are connected by the equation
\[L_{Df}=DL_f.\]

(ii) When $g=Mh$, the Frechet derivative operator of $g$ is
antisymmetric: \[{L_g}^*=-L_g.\]

Evidently the second fact gives a necessary condition for $g$ to
be a gradient. But it is also sufficient:

$\smallskip$

\noindent\textbf{Proposition} Given $g\in \mathcal{A}$. The
antisymmetry equation ${L_g}^*=-L_g$ in $\mathcal{O}_p\mathcal{A}$
is necessary and sufficient for the existence of $h$ with $Mh=g.$

\noindent\textbf{Proof} Suppose first that $g$ is oriented and
satisfies $Eg=ng , n\geq 0 ,$ as well as the hypothesis
${L_g}^*=-L_g.$ If $Pg=g(-1)^\nu ,$ then the Frechet derivative
operator of element $h=a_1g$ is given by
\begin{eqnarray*}L_h&=&(Pg)I+\sum_{n=1}^\infty\left(P^n(-1)^na_1\frac{\partial
g}{\partial a_n}\right)D^{n-1} \\&=&(Pg)I+a_1L_g.\end{eqnarray*}

From $L_{Pg}=-PL_gP$ we see that $L_g$ has orientation opposite to
that of $g$, that is, $PL_gP=L_g(-1)^{\nu+1}.$ This permits the
adjoint of $L_h$ to be calculated as
\begin{eqnarray*}L_h^*&=&(Pg)I+L_g^*(a_1I)(-1)^{\nu+1}
\\&=&(-1)^\nu \left\{gI+L_g(a_1I)\right\}.\end{eqnarray*}

This shows that $(-1)^\nu L_h^*=(g+Eg)+(?)D+\cdots$

But it was seen before that ${L_h}^*=(PMh)I+(?)D+\cdots$ Observing
$PM=-MP$ and $Ph=h(-1)^{\nu+1}$, we conclude that
$M(a_1g)=g+Eg=(n+1)g$ in consequence of three assumptions made at
the beginning of this proof. Returning now to the general case, we
note that operator adjoints, Frechet derivative operators , and
the parity automorphism are interconnected by
\begin{eqnarray*}(PL_gP)^*&=&PL_g^*P
\\&=&L_{Pg}=-PL_gP.\end{eqnarray*}

Thus, if $g\in \mathcal{A}$ has an antisymmetric Frechet
derivative operator then so do $Pg$ and $g\pm Pg$. Hence it
suffices to treat only the case of oriented $g$. Expanding
$g=g_0+g_1+g_2+\cdots$ by homogeneous components in
$\mathcal{A}_0\oplus \mathcal{A}_1\oplus\cdots,$ we observe that
the coefficients of $L_{g_n}$ and $L_{g_n}^*$ fall within
$\mathcal{A}_{n-1}.$ Therefore the antisymmetry of $L_g$ implies
the antisymmetry of all the $L_{g_n}$. From what was said before
$g=Mh$ with $h=a_1(g_0+\frac{1}{2}g_1+\frac{1}{3}g_2+\cdots).$

This completes the proof.

$\smallskip$

\subsection{Summary: the Exact Sequence} A ring $\mathcal{A}$, the
``free SUSY derivation ring on a single fermionic generator", has
been constructed. It has a parity automorphism $P$, a canonical
superderivation $D$, and a SUSY gradient operator $M$.

Necessary and sufficient conditions have been given for
recognizing which elements of $\mathcal{A}$ are derivatives and
which are gradients. In terms of the SUSY gradient operator $M$,
the Frechet derivative operator $L_g$, and the operator adjoint
construction $L\rightarrow L^*,$ these conditions are expressed by
the following exact sequence of rings and linear transformations:

\begin{eqnarray*}\mathcal{O}_p\mathcal{A}\longleftarrow \mathcal{A}\longleftarrow
&\mathcal{A}&\longleftarrow \mathcal{A}\longleftarrow
\mathbb{R}\longleftarrow 0 \\&Df&\longleftarrow f
\\Mh\longleftarrow &h&
\\L_g+L_g^*\longleftarrow g\hspace{7mm}& &\end{eqnarray*}

The sequence is exact in that the kernals of the outgoing
transformations coincide with the images of the incoming
transformations.

\section{The SUSY heat operator }
The ordinary heat equation with potential $u(x)$ and temperature
function $f(x,t)$ is
\[L_uf=0,\hspace{3mm}L_u=\frac{\partial}{\partial
t}-\triangle+u(x).\] Its Green's function evaluated at a field
point $p=(x,t)$ and source point $q=(x^\prime,0)$ may be written
as $G(p,q)=\mathcal{G}_t(x,x^\prime).$ For fixed $x^\prime$ and
variable $x$ and $t$ it satisfies the above heat equation with
initial value
\[\lim_{t\downarrow0}\mathcal{G}_t(x,x^\prime)=\delta(x-x^\prime).\] When the
potential is zero the Green's function is
\[g_t(x-x^\prime)\equiv g(x-x^\prime,t)=\frac{1}{\sqrt{4\pi t}}\exp\left(-\frac{{(x-x^\prime)}^2}{4t}\right).\]
It has the basic properties

\beq\begin{array}{l}i) g_t(x-x^\prime)>0\\

ii) \int_{{\mathbb{R}}^n}g_t(x-x^\prime)dx=1\\

iii)g_{t+s}(x-x^\prime)=\int_{{\mathbb{R}}^n}g_s(y-x^\prime)g_t(x-y)dy.\end{array}\label{proofg}\eeq

$\smallskip$

These properties allow the definition of the conditional Wiener
measure, which may be used to express the Green's function
$\mathcal{G}_t(x,x^\prime)$ of the operator $L_u$

by the Feyman-Kac formula \cite{Glimm}, when the potential $u(x)$
is real and bounded from below.

 For bounded potentials the Green's
function admits an asymptotic expansion
\[\mathcal{G}_t(x,x^\prime)=g(x-x^\prime,t)\sum_{n=0}^\infty
\frac{1}{n!}a_n(x,x^\prime)t^n\] in which the coefficients
$a_n(x,x^\prime)$ are determined recursively by
$a_0(x,x^\prime)=1$,
\[\left(n+(x-x^\prime)\partial_x\right)a_n(x,x^\prime)=\left(\partial_x^2+u(x)\right)a_{n-1}(x,x^\prime).\]
On the diagonal $x=x^\prime$, one has
\[a_n(x,x)=g_n\left(u(x),u^\prime(x),\ldots\right),\] a
finite polynomial in the potential function $u(x)$ and its
derivatives.

Then the equations of the KdV hierarchy \cite{Avramidi}, for
unknown functions $w(x,t)$, are \[w_t=\frac{\partial}{\partial
x}g_n(w,w_x,w_{xx},\ldots)\]

$\smallskip$

We now present a supersymmetric extension of this construction.
The potential and the temperature function now have their values
in an exterior algebra $\Lambda$, also called a Grassmann algebra.

If anticommuting generators of $\Lambda$ are written as
$\theta,\theta_2,\theta_3,\ldots,\theta_m$ then every element of
$\Lambda$ has a unique presentation \[\Phi=\xi+\theta u\] where
$\xi$ and $u$ are in the subalgebra of $\Lambda$ generated by
$\theta_2,\theta_3,\ldots,\theta_m$. Then, defining
$\partial_\theta\Phi=u,$ we obtain a superderivation
$\partial_\theta:\Lambda\rightarrow\Lambda$, that is,
\[\partial_\theta(\Phi_1\Phi_2)=\left(\partial_\theta\Phi_1\right)\Phi_2+
\bar{\Phi}_1\left(\partial_\theta\Phi_2\right)\] where
$\Phi\rightarrow\bar{\Phi}$ is the parity automorphism of
$\Lambda$.

The operator $D=\partial_\theta+\theta\partial_x$ then acts on
``superfields", that is, on differentiable functions $
\mathbb{R}\rightarrow\Lambda$. Using $\Phi_1$ and $\Phi_2$ to
designate superfields, one can check \[D\left(\Phi_1\Phi_2\right)=
\left(D\Phi_1\right)\Phi_2+\bar{\Phi}_1\left(D\Phi_2\right),\]
\[D^2=\partial_x.\] It will be assumed that the potential $\Phi$
is fermionic: $ \bar{\Phi}=-\Phi$ meaning that $ \bar{\xi}=-\xi$
and $ \bar{u}=u$.

$\smallskip$

$\smallskip$

If we take the dimension of $x$ to be $1$: \[[x]=1,\] then one
must have \[[\theta]=\frac{1}{2}\]
\[[D]=-\frac{1}{2},\] consequently
\[[u]=-2,\]
\[[\Phi]=[\xi]=-\frac{3}{2}.\] The most general supersymmetric
extension of $L_u$, assuming positive powers of $D$, becomes then
\[\mathbf{L}=\frac{\partial}{\partial
t}-(D^4-D\Phi+\lambda\Phi D)\] where $\lambda$ is a constant,
dimensionless parameter.

When the superpotential $\Phi$ is zero, it reduces to the heat
operator, while if $\xi=0$ and $\theta=0$ it reduces to $L_u.$

The parameter $\lambda$ already appeared in the analysis of
Mathieu \cite{M} for all supersymmetric extensions of the KdV
equation. The case $\lambda=1$ was related to the integrable
supersymmetric extension of the KdV equation. We will consider in
what follows $\lambda=1.$

There are two supersymmetric extensions of $\delta(x-x^\prime)$.

$\delta(x-x^\prime)\delta(\theta-\theta^\prime)$ and
$\delta(x-x^\prime-\theta\theta^\prime)=\delta(x-x^\prime)-\theta\theta^\prime\delta^\prime(x-x^\prime)$
are both invariants under the supersymmetric transformations
\beq\begin{array}{c}x\rightarrow
x+\theta\eta\hspace{3mm},\theta\rightarrow\theta+\eta \\
x^\prime\rightarrow
x^\prime+\theta^\prime\eta\hspace{3mm},\theta^\prime
\rightarrow\theta^\prime+\eta.\end{array}\label{susytr}\eeq We may
then consider two Green's functions according to each posible
initial conditions. We will denote the corresponding Green's
function by $\mathbf{K}_t(x,x^\prime,\theta,\theta^\prime)$ and $
\mathbf{G}_t(x,x^\prime,\theta,\theta^\prime)$ respectively.

The Green's function for the potential $\Phi$, as a function of
the source point $q=(x^\prime,0)$ and field point $p=(x,t)$, is to
be a function
$\mathbf{K}_t(x,x^\prime,\theta,\theta^\prime),\mathbf{G}_t(x,x^\prime,\theta,\theta^\prime)$
having values in $\Lambda$ and satisfying $
\mathbf{L}\mathbf{K}_t=0,\mathbf{L}\mathbf{G}_t=0$ when $t>0$,
while
$\lim_{t\downarrow0}\mathbf{K}_t=\delta(x-x^\prime)\delta(\theta-\theta^\prime)$,

$\lim_{t\downarrow0}\mathbf{G}_t=\delta(x-x^\prime)-\theta\theta^\prime\delta^\prime(x-x^\prime)$.

$\smallskip$

$\mathbf{K}_t$ and $ \mathbf{G}_t$ are related by
\[-D^\prime \mathbf{K}_t=\mathbf{G}_t,\] in which \[D^\prime=\frac{\partial}{\partial\theta^\prime}+
\theta^\prime\frac{\partial}{\partial x^\prime}\] is the
superderivative with respect to $(x^\prime,\theta^\prime)$.

$\smallskip$

 The Green's function $ \mathbf{K}_t$ may then be expressed as \begin{eqnarray*}&&
\mathbf{K}_t(x,x^\prime;\theta,\theta^\prime)={\mathcal{K}}_t(x-x^\prime
,\theta-\theta^\prime)-{\left\langle\widetilde{D}
\left[\Phi(\tilde{x},\tilde{\theta})\mathcal{K}_{\tilde{t}}(\tilde{x}-x^\prime
,\tilde{\theta}-\theta^\prime)\right]{\mathcal{K}}_{t-\tilde{t}}(x-\tilde{x},\theta
-\tilde{\theta})\right\rangle}_{\tilde{x},\tilde{t},\tilde{\theta}}+  \\
&&+{\left\langle{\left\langle\widetilde{D}\left[\Phi(\tilde{x},\tilde{\theta})
{\mathcal{K}}_{\tilde{t}}(\tilde{x}-x^\prime,\tilde{\theta}-\theta^\prime)\right]
\widetilde{\widetilde{D}}\left[\Phi(\widetilde{\widetilde{x}},\widetilde{\widetilde{\theta
}}){\mathcal{K}}_{\widetilde{{\widetilde{t}}}-\widetilde{t}}(\widetilde{\widetilde{x}}
-\widetilde{x},\widetilde{\widetilde{\theta }}-\widetilde{\theta
})\right]{\mathcal{K}}_{t-\widetilde{\widetilde{t}}}(x-\widetilde{\widetilde{x}},
\theta-\widetilde{\widetilde{\theta}})\right\rangle}_
{\tilde{x},\tilde{t},\tilde{\theta}}\right\rangle}_{\widetilde{\widetilde{x}},
\widetilde{\widetilde{t}},
\widetilde{\widetilde{\theta}}}-
\\&&-\ldots
\end{eqnarray*} where \[\mathcal{K}_t(x-x^\prime,\theta-\theta^\prime)
=g_t(x-x^\prime)(\theta-\theta^\prime),\]
\[0<\widetilde{t}<\tilde{\tilde{t}}<\cdots <t,\]
\[\widetilde{D}=\tilde{\theta}\frac{\partial}{\partial\tilde{x}}+
\frac{\partial}{\partial\tilde{\theta}}\] and
\[{\left\langle\cdot\right\rangle}_{\tilde{x},\tilde{t},\tilde{\theta}}=\int\frac{\partial}
{\partial\tilde{\theta}}(\cdot)|_{\tilde{\theta}=0}d\tilde{x}d\tilde{t}.\]

 The integration on $\tilde{x}$ is over $\mathbb{R}$ , the
integration on the $ \tilde{t}$ variable is according to the above
ordering while the integration on the odd coordinate means
${[\frac{\partial}{\partial\tilde{\theta}}]}_{\tilde{\theta}=0}$
and it must be performed in the order indicated in our formula. We
will show that this series expansion is convergent, under the
assumption of bounded potentials.

The generic term of the expansion may be constructed from the
previous one replacing
$\mathcal{K}_{t-\tilde{\tilde{t}}}(x-\tilde{\tilde{x}},
\theta-\tilde{\tilde{\theta}})$ by
$\hat{D}\left[\Phi(\hat{x},\hat{\theta})\mathcal{K}_
{\hat{t}-\tilde{\tilde{t}}}(\hat{x}
-\tilde{\tilde{x}},\hat{\theta}-\tilde{\tilde{\theta}})
\right]{\mathcal{K}}_{t-\hat{t}}(x-\hat{x}, \theta-\hat{\theta})$
and performing an overall integration on $
\hat{x},\hat{\theta},\hat{t}$ where this point is an intermediate
one with $ \tilde{\tilde{t}}<\hat{t}<t$.

$\smallskip$

The explicit expansion for $\mathbf{G}_t$ turns out to be, after
some calculations,
\begin{eqnarray*}&&
\mathbf{G}_t(x,x^\prime;\theta,\theta^\prime)=g_t(x-x^\prime
-\theta\theta^\prime)-{\left\langle
\Phi(\tilde{x},\tilde{\theta})g_{\tilde{t}}(\tilde{x}-x^\prime
-\tilde{\theta}\theta^\prime)g_{t-\tilde{t}}(x-\tilde{x}-\theta
\tilde{\theta})\right\rangle}_{\tilde{x},\tilde{t},\tilde{\theta}}+  \\
&&+{\left\langle{\left\langle\Phi(\tilde{x},\tilde{\theta})
g_{\tilde{t}}(\tilde{x}-x^\prime-\tilde{\theta}\theta^\prime)
\Phi(\widetilde{\widetilde{x}},\widetilde{\widetilde{\theta
}})g_{\widetilde{{\widetilde{t}}}-\widetilde{t}}(\widetilde{\widetilde{x}}
-\widetilde{x}-\widetilde{\widetilde{\theta }}\widetilde{\theta
})g_{t-\widetilde{\widetilde{t}}}(x-\widetilde{\widetilde{x}}
-\theta\widetilde{\widetilde{\theta}})\right\rangle}_
{\tilde{x},\tilde{t},\tilde{\theta}}\right\rangle}_{\widetilde{\widetilde{x}},\widetilde{\widetilde{t}},
\widetilde{\widetilde{\theta}}}-
\\&&-\ldots
\end{eqnarray*}

The generic term of the expansion may be constructed from the
previous one replacing
$g_{t-\widetilde{\widetilde{t}}}(x-\widetilde{\widetilde{x}}
-\theta\widetilde{\widetilde{\theta}})$ by
$\Phi(\hat{x},\hat{\theta})g_{\hat{t}-\tilde{\tilde{t}}}(\hat{x}
-\tilde{\tilde{x}}-\hat{\theta}\tilde{\tilde{\theta}})
g_{t-\hat{t}}(x-\hat{x} -\theta\hat{\theta})$ and performing an
overall integration on $ \hat{x},\hat{\theta},\hat{t}$ where this
point is an intermediate one with $ \tilde{\tilde{t}}<\hat{t}<t.$

In the bosonic limit, when $\xi(x)=0$ and
$\theta=\theta^\prime=0$, the integration on the odd variables
$\tilde{\theta}$ becomes straightforward and the formula reduces
to
\begin{eqnarray*}&&
\mathbf{G}_t(x,x^\prime;0,0)|_{\xi=0}=g_t(x-x^\prime)-{\left\langle
u(\tilde{x})g_{\tilde{t}}(\tilde{x}-x^\prime)g_{t-\tilde{t}}(x-\tilde{x})\right\rangle}
_{\tilde{x},\tilde{t}}+  \\
&&+{\left\langle{\left\langle u(\tilde{x})
g_{\tilde{t}}(\tilde{x}-x^\prime)
u(\widetilde{\widetilde{x}})g_{\widetilde{{\widetilde{t}}}-\widetilde{t}}(\widetilde{\widetilde{x}}
-\widetilde{x})g_{t-\widetilde{\widetilde{t}}}(x-\widetilde{\widetilde{x}}
)\right\rangle}_
{\tilde{x},\tilde{t}}\right\rangle}_{\widetilde{\widetilde{x}},\widetilde
{\widetilde{t}}}-
\\&&-\ldots
\end{eqnarray*}

This is exactly the Green's function $ \mathcal{G}$ for the
operator $L_u=\partial_t-\triangle+u(x)$.  If we assume $u(x)$ to
be bounded and continuous: \[|u(x)|<M,\] it then follows, using
the semigroup property for the Green's function, that
\[|\mathcal{G}_t(x,x^\prime)|<e^{tM}g_t(x-x^\prime).\] It may also
be shown that the convergent expansion for
$\mathcal{G}_t(x,x^\prime)$ is equal to the Feynman-Kac formula
\[\mathcal{G}_t(x,x^\prime)=\int
dW_t(x,x^\prime)\exp\left(-\int_{-\frac{t}{2}}^{+\frac{t}{2}}u(x(s))ds\right)\]
where $dW_t(x,x^\prime)$ denotes the Wiener measure for the
continuous paths between $x^\prime$ and $x$. When $u(x)$ is real
and bounded both formulas are exactly the same. However, the
Feynman-Kac formula may be established even when $u(x)$ is bounded
from below and $\triangle+u(x)$ is (essentially) self adjoint,
while the expansion in terms of the potential is valid when $u$ is
bounded without assuming the self adjoint property of the
operator.

 $\mathcal{G}_t(x,x^\prime)$ is positive. This property arises directly from the Feynman-Kac
formula.

 It also satisfies
\[\mathcal{G}_t(x,x^\prime)=\mathcal{G}_t(x^\prime,x)\] under the interchange
of the positions of the field point and the source one.

In the next section we will analize these properties for the
supersymmetric extensions we are considering.
\section{ The SUSY Green's function and the SKdV hierarchy } The Green's function
$\mathbf{G}_t$ depends on $x,\theta$ and $x^\prime,\theta^\prime$
and on the components of the superpotential $u$ and $\xi$. In
order to analize the transformation law, under supersymmetry, of
$\mathbf{G}_t$ we will write explicitly its dependence on $u$ and
$\xi$, $\mathbf{G}_t(x,x^\prime;\theta,\theta^\prime;u,\xi).$
Under the supersymmetric transformations :
\begin{eqnarray*}&& x\longrightarrow x+\delta x=x-\eta\theta\\&&
\theta\longrightarrow\theta+\delta\theta=\theta+\eta\\&&
\Phi\longrightarrow\Phi+\delta\Phi=\Phi-\eta
u(x)+\eta\theta\xi^\prime(x),
\end{eqnarray*}

 the components of $\Phi$ transforms  as \begin{eqnarray*}&&
u\longrightarrow u+\delta u=u-\eta\xi^\prime\\&&
\xi\longrightarrow \xi+\delta\xi=\xi-\eta u.
\end{eqnarray*}

 $\mathbf{G}_t$ is then invariant under these
 transformations. That is,
\beq\begin{array}{l}\mathbf{G}_t(x+\delta x,x^\prime+\delta
x^\prime;\theta+\delta\theta,\theta^\prime+\delta\theta^\prime;
u+\delta u,\xi+\delta\xi)=\\=\mathbf{G}_
t(x,x^\prime;\theta,\theta^\prime;u,\xi)\end{array}\label{Ktr}\eeq

To show this invariance property of $\mathbf{G}_t$ we notice that
\[(\Phi+\delta\Phi)
(\tilde{x},\tilde{\theta})=\Phi(\tilde{x}-\delta\tilde{x},\tilde{\theta}-
\delta\tilde{\theta}).\] We then evaluate the left hand member of
(\ref{Ktr}) using the previous expansion formula and perform a
change of variable at each intermediate point
$\tilde{x},\tilde{\theta}$:
\begin{eqnarray*}&& \tilde{x}\longrightarrow
\tilde{x}_1=\tilde{x}-\delta\tilde{x}\\&&
\tilde{\theta}\longrightarrow
\tilde{\theta}_1=\tilde{\theta}-\delta\tilde{\theta},
\end{eqnarray*}
with Jacobian equal to $1$.

We then use the property that the combination
$(\tilde{\tilde{x}}-\tilde{x}-\tilde{\tilde{\theta}}\tilde{\theta})$
is invariant under this change of coordinates. We end up with the
relation (\ref{Ktr}).

The other symmetry of the Susy Green's function $\mathbf{G}_t$ is
\beq\mathbf{G}_t(x,x^\prime;\theta,\theta^\prime;u,\xi)
=\mathbf{G}_t(x,x^\prime;\theta^\prime,\theta;u,\xi).\eeq It
follows by performing changes of variables on the time arguments
at each integrand on each term of the expansion. In terms of the
components of $ \mathbf{G}_t$ it means
\[\mathbf{G}_t(x,x^\prime;\theta,\theta^\prime)=A_t(x,x^\prime)
+\theta^\prime B_t(x,x^\prime)+\theta
C_t(x,x^\prime)+\theta\theta^\prime D_t(x,x^\prime)\]
\[A_t(x,x^\prime)=A_t(x^\prime,x)\]\[B_t(x,x^\prime)=C_t(x^\prime,x)\]
\[D_t(x,x^\prime)=-D_t(x^\prime,x).\]

We will now evaluate $\mathbf{G}_t$ by performing all integrations
on the odd variables.  We start evaluating
$\mathbf{G}_t(x,x^\prime;0,0;u,\xi)$. We denote
\[\xymatrix{\tilde{x},\tilde{t} \ar@{=>}[r]  & \tilde{\tilde{x}},
\tilde{\tilde{t}}}=\mathcal{G}_{\tilde{\tilde{t}}-\tilde{t}}
(\tilde{\tilde{x}},\tilde{x}) \hspace{3mm}\mathrm{\:the\:}
\hspace{0.5mm}\mathrm{\:bosonic\:}\hspace{0.5mm}\mathrm{\:propagator\:},
\]

\[ \xymatrix{\tilde{x},\tilde{t} \ar@{~>}[r]  & \tilde{\tilde{x}},
\tilde{\tilde{t}}}=-\frac{1}{2}\frac{(\tilde{\tilde{x}}-
\tilde{x})}{(\tilde{\tilde{t}}-\tilde{t})}g_{\tilde{\tilde{t}}-\tilde{t}}
(\tilde{\tilde{x}}-\tilde{x}) \hspace{3mm}\mathrm{\:the\:}
\hspace{0.5mm}\mathrm{\:fermionic\:}\hspace{0.5mm}\mathrm{\:propagator\:}.\]

 An arrow followed by a vertex
$\xi(\tilde{\tilde{x}})$ denotes multiplication of the propagator
 by the vertex and integration on the corresponding coordinates
$\tilde{\tilde{x}},\tilde{\tilde{t}}.$

The Green's function at $\theta=\theta^\prime=0$ may then be
expressed by

$\smallskip$

\begin{eqnarray*}&&\mathbf{G}_t(x,x^\prime;0,0;u,\xi)=\hspace{3mm}\xymatrix{x^\prime,t^\prime
\ar@{=>}[r] &
x,t\hspace{3mm}+\hspace{3mm}x^\prime,t^\prime\ar@{=>}[r] &
 \xi\ar@{~>}[r] & \xi\ar@{=>}[r] & x,t \hspace{3mm}+ }\\
 &&\xymatrix{+ \hspace{3mm}
 x^\prime,t^\prime
\ar@{=>}[r] & \xi\ar@{~>}[r] & \xi\ar@{=>}[r] & \xi\ar@{~>}[r] &
\xi \ar@{=>}[r]& x,t \hspace{3mm}+}\\
&&\xymatrix{+ \hspace{3mm} x^\prime,t^\prime \ar@{=>}[r] &
\xi\ar@{~>}[r] & \xi\ar@{=>}[r] & \xi\ar@{~>}[r] & \xi
\ar@{=>}[r]& \xi\ar@{~>}[r] & \xi\ar@{=>}[r] & x,t
\hspace{3mm}+\cdots}\end{eqnarray*}

$\smallskip$

It can be shown that this expansion on the fermionic vertex $\xi$
is convergent provided $u(x)$ is bounded and $\xi(x)$ is bounded
in the following sense. It is possible to express the product
\beq\xi( \tilde{\tilde{x}})\xi(
\tilde{\tilde{x}})=\frac{1}{2}(x-\tilde{\tilde{x}})f(
\tilde{\tilde{x}},\tilde{x})\label{hip}\eeq since the left hand
member is antisymmetric on $ \tilde{x}\leftrightarrow
\tilde{\tilde{x}}.$ We assume then \beq\begin{array}{l}|u(x)|<M,
\\f( \tilde{\tilde{x}},\tilde{x})<M^2.\end{array}\label{cotas}\eeq The square
arises from dimensional arguments. In fact, let us remember that $
[\xi]=-\frac{3}{2}$ and $[u]=-2$ and hence $[f]$ must be $-4$.
After replacing (\ref{hip}) in the expression of
$\mathbf{G}_t(x,x^\prime;0,0;u,\xi)$, the contributions of the
fermionic propagator times the fermionic vertices may be work out
in terms of derivatives of $g_t$. One may then use (\ref{cotas})
and the semigroup properties for $g_t$ to obtain a bound for the
series expansion of $\mathbf{G}_t(x,x^\prime;0,0;u,\xi)$.

The complete expression for
$\mathbf{G}_t(x,x^\prime;\theta,\theta^\prime)$, expressed in
terms of its value at $\theta=\theta^\prime=0$, is the following

\begin{eqnarray*}&&{\mathbf{G}}_t(x,x^\prime;\theta,\theta^\prime)=\mathbf{G}_t
(x,x^\prime,0,0)-\theta{\left\langle \mathbf
{G}_{\tilde{t}-t^\prime}(\tilde{x},x^\prime,0,0)
 \xi(
 \tilde{x})g_{t-\tilde{t}}^\prime(x-\tilde{x})\right\rangle}_{\tilde{x},\tilde{t}}+\\
&&+\theta^\prime{\left\langle\xi( \tilde{x})
g_{\tilde{t}-t^\prime}^\prime(\tilde{x}-x^\prime) {\mathbf
{G}}_{t-\tilde{t}}(x,\tilde{x},0,0)\right\rangle}_{\tilde{x},\tilde{t}}-\\
 &&-\theta\theta^\prime{\left\langle{\left\langle
\xi( \tilde{x})g_{\tilde{t}-t^\prime}^\prime(\tilde{x}-x^\prime)
{\mathbf
{G}}_{\widetilde{{\widetilde{t}}}-\widetilde{t}}(\widetilde{\widetilde{x}},
\widetilde{x};0,0)\xi( \tilde{\tilde{x}})
g_{t-\widetilde{\widetilde{t}}}^\prime(x-\widetilde{\widetilde{x}})\right\rangle}_
{\tilde{x},\tilde{t}}\right\rangle}_{\widetilde{\widetilde{x}},\widetilde{\widetilde{t}}
}\end{eqnarray*}

Just as in the bosonic case, $\mathbf{G}_t$ possesses an
asymptotic expansion
\[\mathbf{G}_t(x,x^\prime,\theta,\theta^\prime)=g_t(x-x^\prime-\theta\theta^\prime)
\sum_{k=0}^\infty
\frac{t^k}{k!}\Gamma_k(x,x^\prime)\] in which each term has the
form
\[\Gamma_k(x,x^\prime)=A_k(x,x^\prime)+\theta B_k(x,x^\prime)+\theta^\prime
C_k(x,x^\prime)+\theta\theta^\prime D_k(x,x^\prime).\] The
approximation of $\mathbf{G}_t$ by its asymptotic expansion proves
that $A_k,B_k,C_k,D_k$ have the same $x,x^\prime$ symmetry as
noted before. As before, $\Gamma_k$ is constructed from the
potential $\Phi(x)=\xi(x)+\theta u(x)$ by an iterative procedure
starting with $\Gamma_0(x,x^\prime)=1$.

Formally equating $x$ with $x^\prime$ and $\theta$ with
$\theta^\prime$ we define
\[g_k(x)=A_k(x,x)+2\theta B(x,x).\]
This must be a polynomial in $\xi(x)$, $u(x)$, and their
derivatives. But it turns out to be expressible as a polynomial in
$\Phi=\xi+\theta u,D\Phi=u+\theta\partial_x\xi,D^2\Phi=
\partial_x\xi+\theta\partial_xu,\ldots$

These polynomials can be seen as elements of the free
supersymmetric derivation ring on a single fermionic generator,
but in the notation $\Phi=a_1,D\Phi=a_2,\ldots$

The first few such polynomials are
\[g_2=-a_2\]\[g_6=-a_6+3a_2^2-2a_1a_3\]\[g_{10}={\left(-a_{10}+10a_2a_6+5a_4^2-10a_2\right)}^3
-(4a_1a_7+a_3a_5-15a_1a_2a_3).\] They have been shown to be
gradients of conserved quantities of the SKdV \cite{M} equation,
whose unknown function $\Omega(x,t)$ is to satisfy \[\Omega_t=
-D^6\Omega+3\Omega\left(D^3\Omega\right)+3\left((D\Omega)(D^2\Omega)\right).\]
The symmetry of the asymptotic Green's function permits the
derivation, after several calculations, of the recursive algorithm
\[D^2g_{n+4}=\left(D^6+2a_1D^3-4a_2D^2+a_3D-2a_4I\right)g_n-a_1l_n,\]
\[D^2l_n=-a_2Dg_n+a_3g_n.\] The members of the super KdV hierarchy
are then given by $Mg_n$ with \[M=D^5-3a_1D^2-a_2D-2a_3I.\] In
particular $Mg_2$ is the super KdV equation of Mathieu \cite{M}
the same one appearing above.

One may use necessary and sufficient conditions arising from the
exact sequence established in section (2) to show $g_n$ are the
gradients of conserved quantities of the SKdV hierarchy.
\section{Conclusions} We introduced the free $N=1$ supersymmetric derivation
ring. We established the exact sequence of supersymmetric rings
and linear
transformations:\begin{eqnarray*}\mathcal{O}_p\mathcal{A}\longleftarrow
\mathcal{A}\longleftarrow &\mathcal{A}&\longleftarrow
\mathcal{A}\longleftarrow \mathbb{R}\longleftarrow 0
\\&Df&\longleftarrow f
\\Mh\longleftarrow &h&
\\L_g+L_g^*\longleftarrow g\hspace{7mm}& &\end{eqnarray*}

Several of these relations were already known in the literature
\cite{Manin,MM}.

 We used necessary and sufficient conditions arising from
this exact sequence to obtain the essential relations between
conserved quantities, gradients and the $N=1$ super KdV
hierarchy. We combine these algebraic conditions together with an
analytical analysis of the super heat operator \[
\mathbf{L}=\frac{\partial}{\partial t}-( D^4-D\Phi+\Phi D).\] We
found an explicit series expansion for the Green's function of the
Super heat operator and discussed its properties. The expansion is
convergent under the assumption of bounded bosonic and fermionic
potentials as established in section (4). This analysis may be
relevant since there is no rigorous Feynman- Kac formula for the
fermionic case.

Finally we show that the asymtotic expansion when
$t\rightarrow0^+$ of the Green's function of $ \mathbf{L}$,
evaluated over its diagonal, generates all the members of the
$N=1$ super KdV hierarchy.

The exact sequence of susy rings and linear transformations may
also be constructed with $N>1$ susy generators, we hope to discuss
this extension elsewhere. We expect that the algebraic
construction established in the first part of this work may have a
natural extension for more general finite-dimensional quantum
systems. This would be of great help in understanding, for
example, the quantum behavior of the supermembrane and super
$D$-brane theories \cite{Mari, Boulton, Alvaro}.

\end{document}